\documentclass{aa}

\usepackage{epsfig}
\newcommand{\Lya}{Ly$\alpha$}
\newcommand{\gsim}{\raisebox{-5pt}{$\;\stackrel{\textstyle >}{\sim}\;$}}
\newcommand{\lsim}{\raisebox{-5pt}{$\;\stackrel{\textstyle <}{\sim}\;$}}

\begin{document}
 
   \thesaurus{03         % A&A Section 3: extragalactic Astronomy
	      (11.09.3;  % intergalactic medium	
               11.17.1;  % quasars: absorption lines
               11.17.4 )} % quasars: individual(Q0307-195A,B) 

    \title{The size and geometry of the \Lya\ clouds 
   	\thanks{Based on observations collected at the European
Southern Observatory, La Silla, Chile (ESO No. 149.B-0013).} 
	}

   \author{V. D'Odorico\inst{1}
	\and S. Cristiani\inst{2}
	\and S. D'Odorico\inst{3}
	\and A. Fontana\inst{4}
	\and E. Giallongo\inst{4}
	\and P. Shaver\inst{3}
          }
 
   \offprints{Stefano Cristiani}
 
   \institute{International School for Advanced Studies, SISSA, via Beirut 
	      2-4, I-34014 Trieste, Italy
	\and Dipartimento di Astronomia dell'Universit\`{a} di Padova,
	      Vicolo dell'Osservatorio 5,\\I-35122 Padova, Italy
	\and European Southern Observatory, Karl Schwarzschild Strasse 2, 
	     D-85748 Garching, Germany
	\and Osservatorio Astronomico di Roma, via dell'Osservatorio, I-00040
	     Monteporzio, Italy
	}
 
%%%*** completa il Received
   \date{Received April 16, 1998; accepted \dots}
 
   \maketitle
 
   \begin{abstract}

Spectra of the QSO pair Q0307-195A,B have been obtained in the \Lya\
forest (3660--3930 \AA) and \ion{C}{iv} (4720--4850 \AA) regions with
a FWHM resolution between 0.7 and 0.5 \AA.
46 lines have been detected in the spectrum of object A 
while 36 in
the spectrum of object B, of them 29 and 20 were identified as \Lya\
absorptions respectively. 
The present observations have been
supplemented with data of comparable quality on other 7 QSO pairs
available in the literature to give an enlarged sample of 217 
\Lya\ lines with rest equivalent width  $W_o\ge 0.3$ \AA. 
The analysis of the hits
%%%***
(i.e. when an absorption line appears in both QSO spectra)
and misses (i.e. when a line is seen in any of the 
QSO spectra, but no line is seen in the other),
carried out with an improved statistical approach,
indicates that the absorbers have typically a large
size\footnote{Throughout this paper we will assume:
$h \equiv H_0 / (100$ km s$^{-1}$ Mpc$^{-1}$), $q_0 = 0.5$.}:  
$R = 362\ h^{-1}$ kpc, with 95\% confidence 
limits $298<R<426\ h^{-1}$ kpc 
and $R = 412\ h^{-1}$ kpc, with 95\% confidence limits 
$333<R<514\ h^{-1}$ kpc
%%%***
for the radius of idealized spherical and disc geometries, respectively. 
The present data do not allow to establish any correlation of the 
typical inferred size with the proper separation or with the redshift of 
the pairs. 

The correlation between the observed equivalent widths of the
absorbers in the adjacent lines of sight becomes poorer and poorer with
increasing proper separation.  A disc geometry with a column density
profile $N(r) \propto (r/R_0)^{-\gamma}$, $\gamma=4$, is found to
reasonably reproduce the data with $R_0\simeq 100-200\ h^{-1}$ kpc, but
also spherical clouds with the same column density profile and a
power-law distribution of radii may give a satisfactory representation
of the observations.

   \end{abstract}
 
%
%________________________________________________________________
 
\section{Introduction}
 
Closely separated QSO pairs and gravitationally lensed QSOs 
provide two or more adjacent lines of sight (LOS), which allow to 
sample the size and clustering of the absorbers.

The ray path separation of gravitationally lensed QSOs usually spans
subgalactic scales; studies of these spectra have shown 
that the \Lya\ clouds are much larger than a few kiloparsecs 
(Weymann \& Foltz 1983; Foltz et al. 1984; Smette et al. 1992, 1995).
Further results again pointing at very large sizes have 
been obtained for LOS separations of a few arcminutes or less 
(Shaver \& Robertson 1983, SR83; Crotts 1989;  
Dinshaw et al. 1995; Dinshaw et al. 1997, D97; Fang et al. 1996, FDCB; 
Crotts \& Fang 1997, CF97).
The first indication in this sense has been obtained 
observing the QSO pair Q1343+2640A,B 
(with a proper sightline separation of $39-40\ h^{-1}$ 
kpc at $z\sim2$), indicating a cloud radius of $100-200\ h^{-1}$ kpc 
(Bechtold et al. 1994; Dinshaw et al. 1994).
QSO pairs with separations larger than a few arcminutes, 
corresponding to proper separations up to  
megaparsecs, have the potential to study the large-scale 
spatial distribution of \Lya\ or metal absorbers (Crotts 1985; 
Jakobsen et al. 1986; Tytler, Sandoval \& Fan 1993; Elowitz, 
Green \& Impey 1995; Dinshaw\& Impey 1996; Williger et al. 1996).

Multiple sightlines are also highly valuable in probing
%%%***
the large scale 
structure. In particular, multiple, well-sampled sightlines can provide a 
stronger test for voids by placing more absorbers in a void-sized volume 
than could possibly be obtained along a single sightline.  

Furthermore, the effect of the UV radiation field of a foreground QSO  
on the absorption spectrum of a background QSO provides an important 
way of testing the 
``proximity effect'' interpretation (Bajtlik, Duncan \& Ostriker 1988) 
of the ``inverse effect'' observed in single QSO sightlines.

\vskip 12pt
   
We present new observations of the QSO pair Q0307-195A,B 
discovered in an objective prism survey by MacAlpine and 
Feldman (\cite{macalp}) (A = \object{UM 680} and B = \object{UM 681}).
Later on, the pair was observed by Shaver and Robertson (\cite{sr83}) 
who analyzed the absorption spectra of the two objects. 
Both QSOs are about of 19th magnitude, their redshifts are $z_\mathrm{em}
(\mathrm{A}) = 2.1439\pm 0.0003$ and $z_\mathrm{em}(\mathrm{B}) = 2.1217\pm 
0.0003$ (SR83), and they are separated by 56 arcseconds on the plane of the 
sky. 
The redshift difference is 
%%%***
small enough ($\sim 2126$ km s$^{-1}$) that they 
could be members of the same cluster.  

The rest of the paper is organized as follows: in \S~2, we present 
spectroscopic observations of Q0307-195A,B and describe the procedure of 
reduction and calibration of the spectra. 
In {\S}~3 the metallic systems in the two QSO spectra are listed and briefly 
discussed. 
Section~4 describes the procedure to obtain an estimate for the radius of 
the \Lya\ absorbers in the hypothesis of simple geometry. In subsection 
4.1 we tackle the same issue on the 
%%%***
basis of more realistic models for the 
absorbers.
Finally, in \S~5 we summarize our results. 

\section {Observations and data reduction}

\begin{figure*}
\epsfxsize=17cm
\epsffile{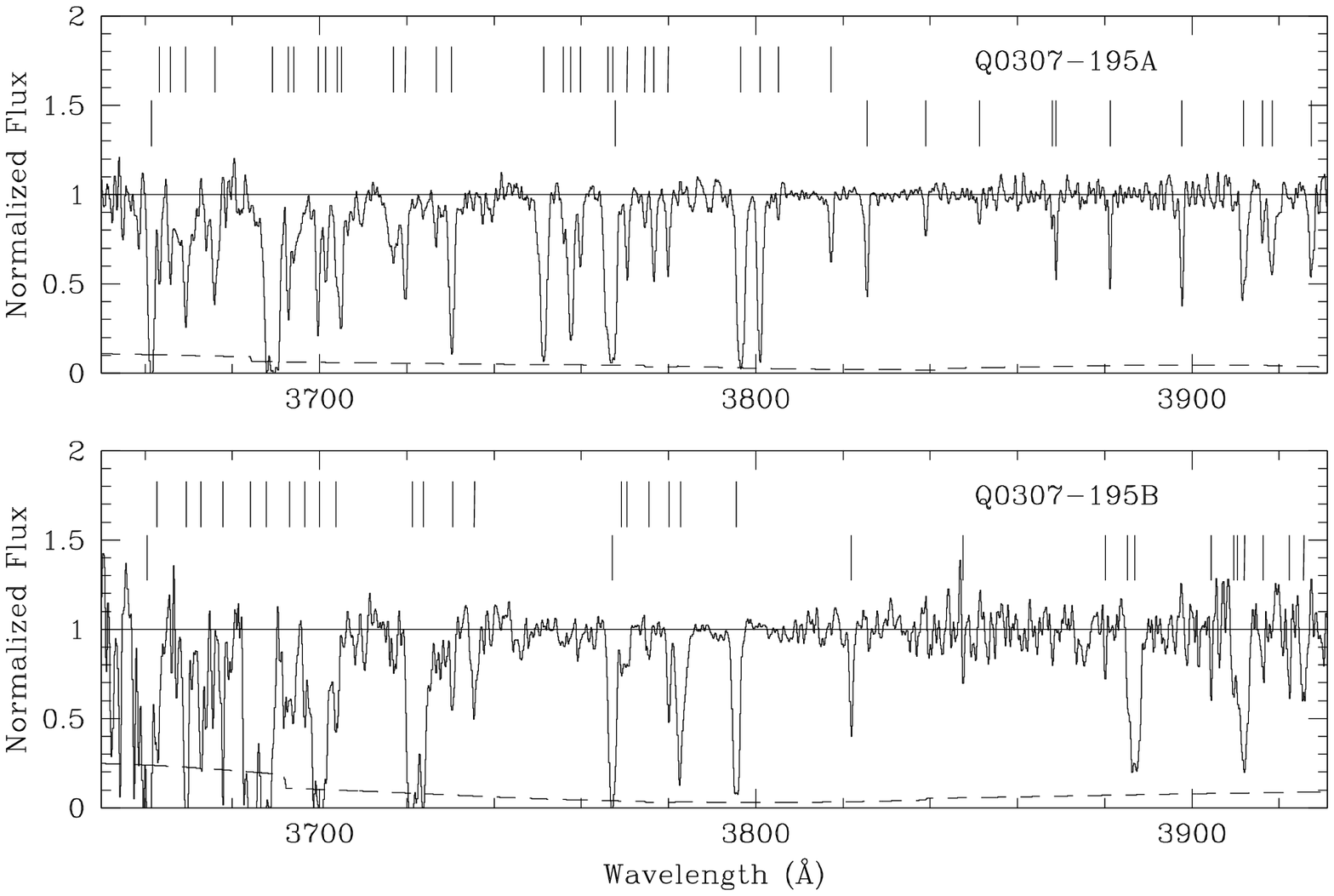}
\vskip -6cm
\caption[]{\label{spec1}Spectra of Q0307-195A,B obtained with the NTT EMMI 
instrument as a function of vacuum heliocentric wavelength. The dashed line 
shows the $1\sigma$ error in the flux. Tick marks indicate absorption lines 
detected at or above the $3.5\sigma$ confidence level, upper and lower 
tick marks are referred to \Lya\ and  metal absorption lines respectively}
\end{figure*}

\begin{figure*}
\epsfxsize=17cm
\epsffile{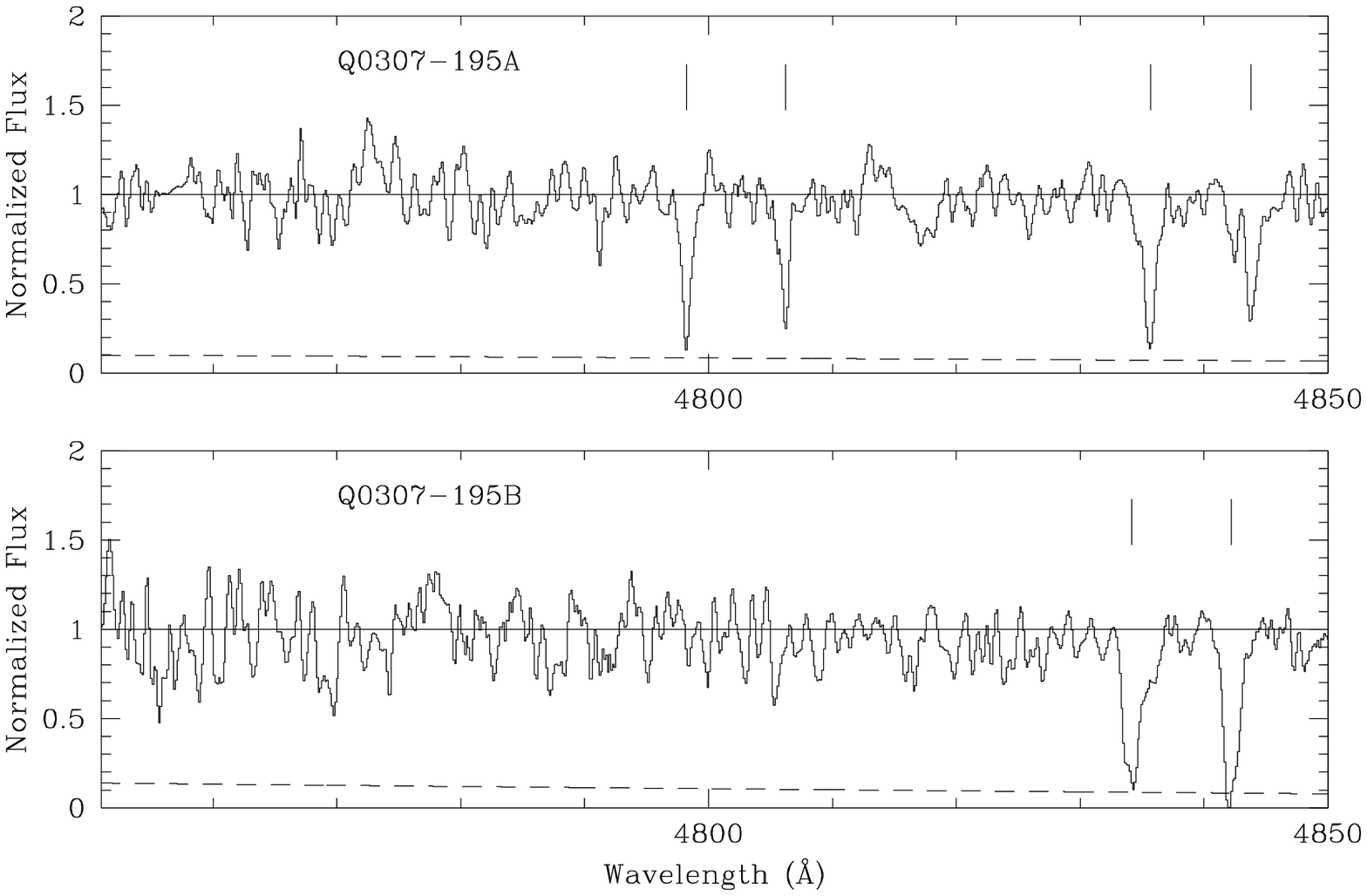}
\vskip -6cm
\caption[]{\label{spec2}Same as Fig.~\ref{spec1}, for the \ion{C}{iv}
region}
\end{figure*}

We obtained spectra of the QSO pair Q0307-195A,B with the blue arm
of the EMMI spectrograph at the 3.5m New Technology Telescope of the 
European Southern Observatory in two runs in November 1994 and 
November 1995. 
The spectrograph was used with the holographic 3000 lines mm$^{-1}$ 
grating (ESO \# 11) and a $1024^{2}$, 24 $\mu$m pixel, Tektronix CCD 
(ESO \# 31). 
Two or three spectra were obtained at each of three different 
central wavelengths: 3690,3750 and 3850 \AA\ (\Lya\ forest region) 
and one at 4780 \AA\ (\ion{C}{iv} region). 
Typical exposure times were 6000 s/spectrum for a total exposure time 
of 56\,900 s. In this instrument configuration one CCD pixel
corresponds on average to 0.15 and  0.13 \AA\ in the UV and blue
regions respectively.
The typical slit width was 1.2 arcsec in width and 180 arcsec in
length and it was aligned at  P.A. $135\degr.6$ to capture both QSOs. 

The spectra were extracted, wavelength calibrated, normalized and merged
using standard procedures of the ESO MIDAS {\em long slit} reduction
package.

The final smoothed and  merged spectra  cover the
ranges $\lambda\lambda 3660-3930$ \AA\ and 
$\lambda\lambda 4720-4850$ \AA. 

The resolution, as measured from the Thorium lines of the
calibration spectra, extracted and treated in the same way 
as the QSOs spectra, is 0.7 and 0.5 \AA\ in the UV and blue 
range respectively. 
The $s/n$ ratio of the final spectra varies from 25 at the peak 
of the \Lya\ emissions of the two QSOs to 5 at the the shortest wavelengths.  
The relative accuracy of the wavelength scales in the spectra of 
the two QSOs is better than 20 km s$^{-1}$.

The normalized spectra are shown in Figs.~\ref{spec1},~\ref{spec2}. 

\begin{table*}
\caption[t1]{Absorption lines in the spectrum of Q0307-195A}
{\label{t1}}
\epsfysize=26cm
\epsffile{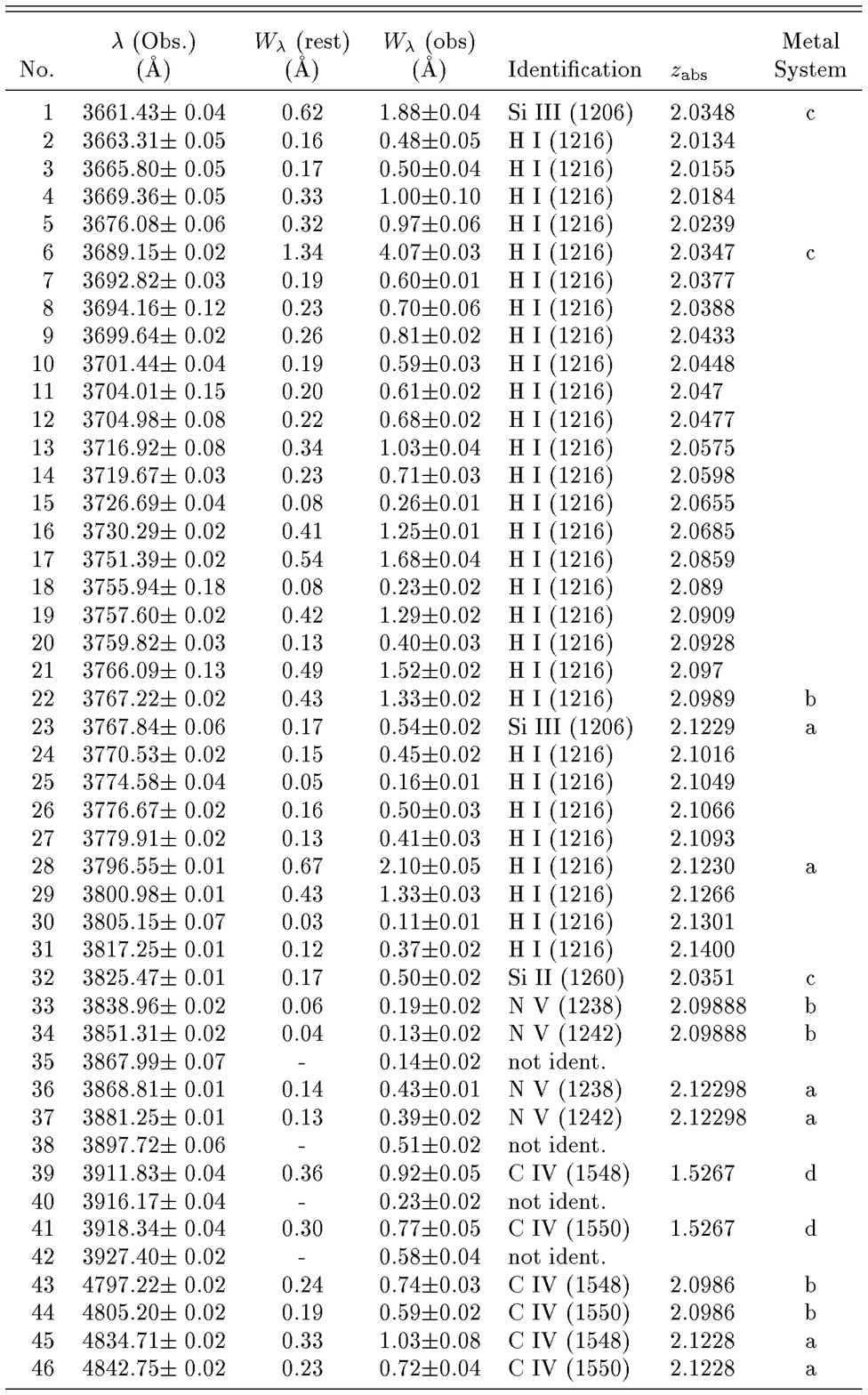}
\end{table*}

\begin{table*}
\caption[t2]{Absorption lines in the spectrum of Q0307-195B}
{\label{t2}}
\epsfysize=26cm
\epsffile{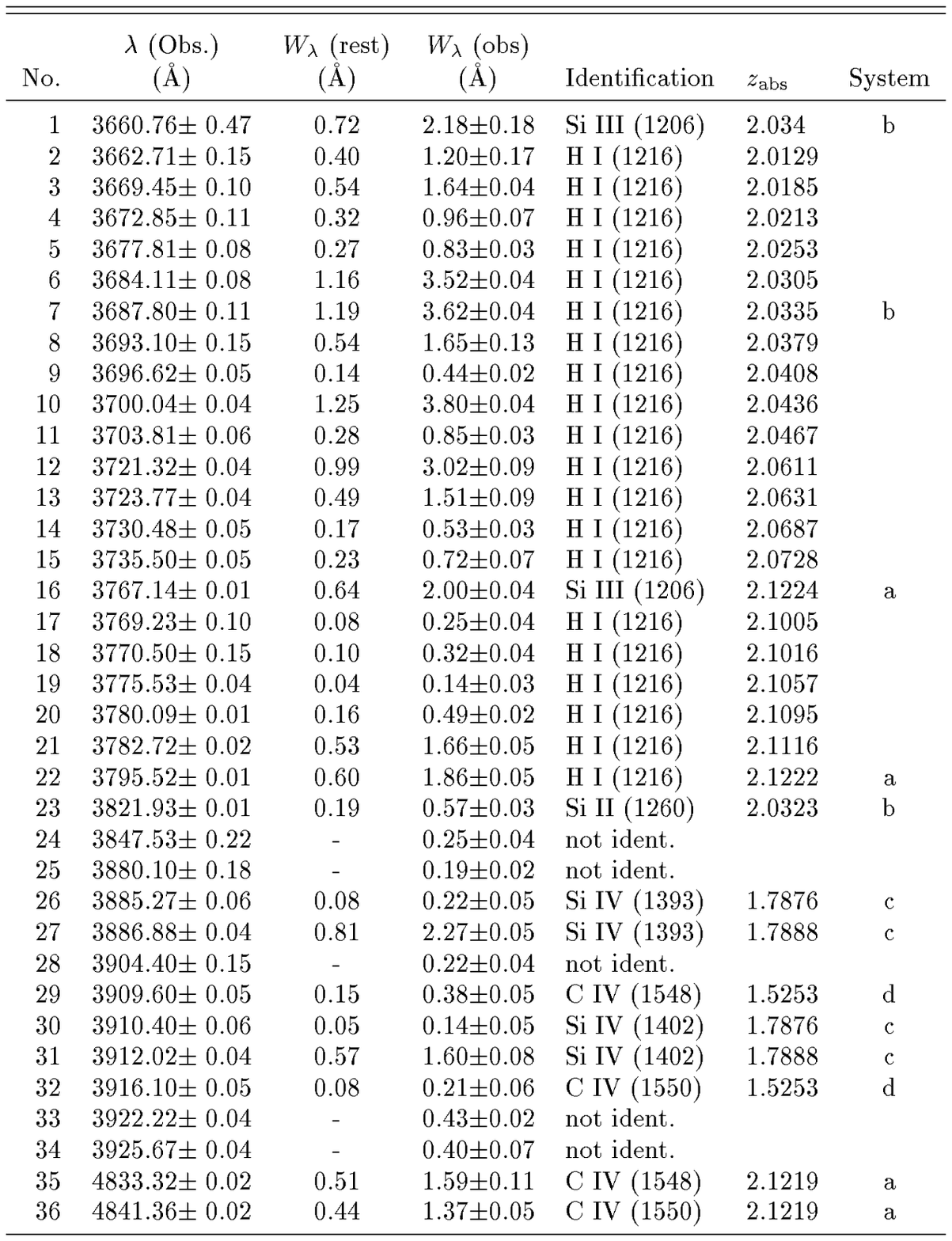}
\end{table*}

\section{Detection and identification of lines}

%%%%%%%% added part %%%%%%%%%%%%%%%%%%%%%%%%%%%%%%%%%%%%%%%%%%%%%%%%%%%%%
The determination of the continuum in the QSO spectrum is a critical step
because it affects the measurement of the absorption line parameters. 
In the present work, we selected the
portions of the spectrum free of strong absorption
lines or artificial peaks (e.g. due to cosmic rays), i.e. where the RMS
fluctuation about the mean becomes consistent with noise statistics, 
and then, we estimated the continuum level by spline-fitting these regions
with quadratic polynomials.
%%%%%%%%%%%%%%%%%%%%%%%%%%%%%%%%%%%%%%%%%%%%%%%%%%%%%%%%%%%%%%%%%%%%%%%%

In order to detect the absorption lines we have searched for all 
the features deviating from the continuum more than 3.5 times the RMS 
noise. 
%%%%%%%% added part %%%%%%%%%%%%%%%%%%%%%%%%%%%%%%%%%%%%%%%%%%%%%%%%%%%%%
The lines have been fitted with Voigt profiles
convolved with the instrumental spread function, making use of a
minimization method of $\chi^2$. This step has been performed within the   
ESO MIDAS {\em lyman} package (Fontana \& Ballester 1995).
The values of the wavelength, redshift $z$, and rest equivalent width
$W_o$ have been determined for isolated lines and individual
components of blends.

The number of components of each absorption feature is assumed to be the
minimum required to give a reduced $\chi^2 <1$  (corresponding to a
confidence level $P\gsim 50 \%$).
%%%%%%%%%%%%%%%%%%%%%%%%%%%%%%%%%%%%%%%%%%%%%%%%%%%%%%%%%%%%%%%%%%%%%%%%%%

%%%%%%%%%%%%%%%%%%%%%%%%%added part%%%%%%%%%%%%%%%%%%%%%%%%%%%%
The error on the observed equivalent widths was computed in such a way
that it includes also the uncertainty in the determination of the
level of the continuum and in the choice of the fitting interval for 
the lines. 
%%%%%%%%%%%%%%%%%%%%%%%%%%%%%%%%%%%%%%%%%%%%%%%%%%%%%%%%%%%%%%%%

Due to the improved signal to noise ratio ($s/n$) and resolution, we 
observe more than twice the number of lines found by SR83 
in the common wavelength interval. The lists of lines are presented in
the Tables~\ref{t1}, \ref{t2}. 

\subsection{Metal Systems}

We based the identification of metal lines  on SR83. 
All the systems reported 
by them in the common wavelength interval are confirmed, 
in addition we found one new system in  Q0307-195A and one 
possible new system in Q0307-195B (see 3.2 and 3.3). 

An identification programme, based on the method of 
Young et al. (1979), has been applied to the newly observed lines 
that could not be attributed to systems already identified by 
SR83. 
Because of the limited wavelength coverage in the red region of the 
spectra, we were able to identify only one new \ion{C}{iv} doublet.

Four and five lines redward of the \Lya\ remained unidentified 
in Q0307-195A and B respectively.

The metal systems of each object are described in detail 
in the following section.

\subsection{Object Q0307-195A}

\begin{description} 
\item[]{\em The metal system at $z_\mathrm{abs} = 2.1229$ - a}

Already observed by SR83. 
This absorption system presents a redshift difference of only 
115 km s$^{-1}$ with object B and of 77 km s$^{-1}$ with system 
{\em a} of object B. 
SR83 argued that this common absorption is likely due to a very 
extended gaseous halo or disc, physically associated with Q0307-195B 
itself. 

\item[]{\em The metal system at $z_\mathrm{abs} = 2.0988$ - b}
 
This is a new system. 
We identified the \ion{C}{iv} and \ion{N}{v} doublets in the red 
portion of the spectrum,  
and the \Lya\ line blended with the \ion{Si}{iii} 1207 of the previous system. 

%%%***
\item[]{\em The metal systems at $z_\mathrm{abs} = 2.03497$ - c
and $z_\mathrm{abs} = 1.5267$ - d}

Already observed by SR83.
No new line has been added.

\end{description} 

\subsection{Object Q0307-195B}

\begin{description} 
%%%***
\item[]{\em The metal systems at $z_\mathrm{abs} = 2.1221$ - a
and $z_\mathrm{abs} = 2.0333$ - b}

Already observed by SR83.
No new line has been added.
 
\item[]{\em The metal system at $z_\mathrm{abs} = 1.7882$ - c}

Already observed by SR83.
The identification is based only on the \ion{Si}{iv} doublet, 
in the \Lya\ forest a possible \ion{C}{ii} can be identified 
but it is blended with a strong \Lya\ line.
The \ion{Si}{iv} doublet shows velocity structure and it 
was fitted with two components. 

\item[]{\em The metal system at $z_\mathrm{abs} = 1.5253$ - d}

The identification of this possible new system is based only 
on the \ion{C}{iv} doublet found redward of the \Lya\ emission line. 
This system differs in redshift by only 166 km s$^{-1}$ 
from system {\em d} in object A. 
\end{description} 
 
All the lines in the \Lya\ forest not identified as metals 
are assumed to be \Lya.  

\section{The size of the \Lya\ absorbers}

\begin{table*}
\caption[thit]{\Lya\ lines that give rise to hits in the spectrum of 
Q0307-195A,B. }
{\label{thit}}
\vskip -9cm
\epsfxsize=17cm
\epsffile{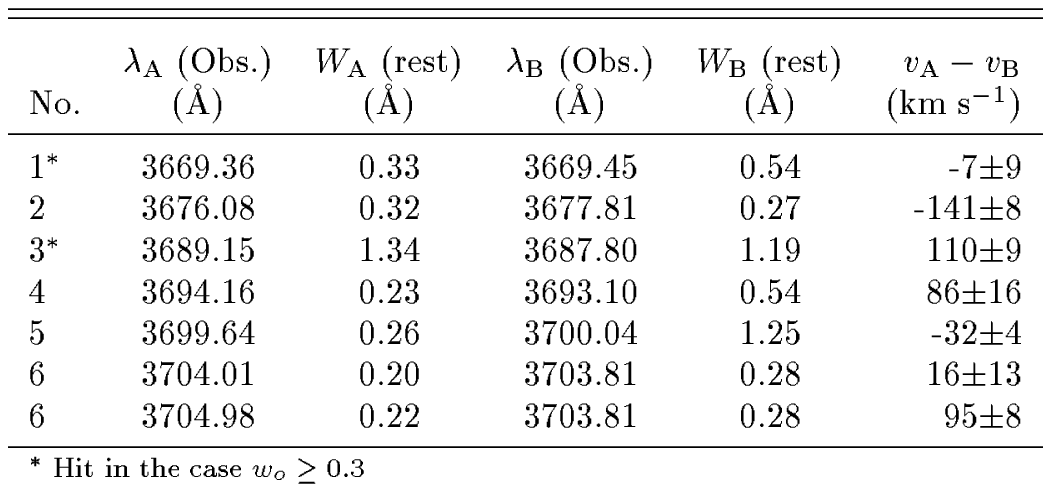}
\vskip -11.5cm
\end{table*}

\begin{table*}
\caption[t3]{Parameters of the QSO pairs forming our enlarged sample and 
\Lya\ cloud radius estimates in the hypothesis of spherical and disc 
geometry}
{\label{t3}}
\vskip -1cm
\epsfxsize=15.2cm
\epsffile{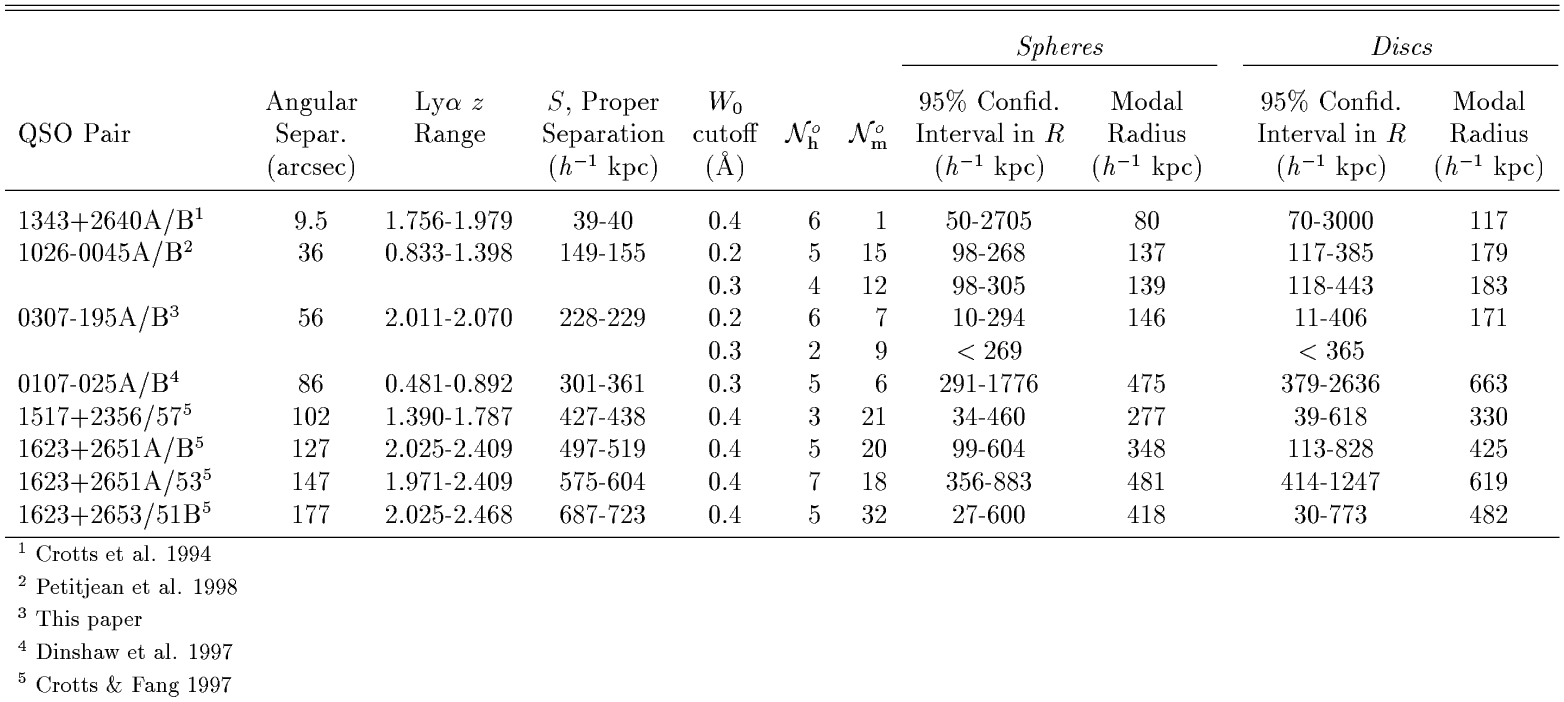}
\vskip -12.5cm
\end{table*}

The main aim of this work is to determine the transverse size of the \Lya\ 
absorbers using the \Lya\ forests of the two adjacent LOSs of the pair.
%%%%%%%%%% added part %%%%%%%%%%%%%%%%%%%%%%%%%%%%%%%%%%%%%%%%%%%%%
In order to avoid spurious effects due to the ``proximity effect'',  
only the \Lya\ lines with velocity separation greater
than 5000 km s$^{-1}$ from the lower QSO redshift in the pair were
considered in the sample.  
%%%%%%%%%%%%%%%%%%%%%%%%%%%%%%%%%%%%%%%%%%%%%%%%%%%%%%%%%%%%%%%%%%%

The spectra of the observed pairs show a number of {\em hits}, 
${\cal N}_\mathrm{h}^o$, and of {\em misses}, ${\cal N}_\mathrm{m}^o$. 
A {\em hit} occurs when an absorption line above a given threshold in rest 
equivalent width, $W_o$, appears in both QSO spectra, 
with a velocity difference less than a limit, $\Delta v$.
A {\em miss} occurs when a line is seen in any of the 
QSO spectra, but no line above the $W_o$ threshold is seen in the other. 

A value $\Delta v = 200$ km s$^{-1}$ has been adopted as the 
most suitable velocity window,  
being significantly larger than the resolution ($\sim 40$ km s$^{-1}$) 
and close to the clustering scale observed for \Lya\ absorbers 
(Cristiani et al. 1997).
%%%%%%%%%% added part %%%%%%%%%%%%%%%%%%%%%%%%%%%%%%%%%%%%%%%%%%%%%
As a consequence, all the \Lya\ lines in the sample whose separation
was less than $200$ km s$^{-1}$ have been merged into a single line 
with wavelength given by the average of the wavelengths 
and equivalent width equal to the sum of the equivalent
widths of the ``parent'' lines. 
%%%%%%%%%%%%%%%%%%%%%%%%%%%%%%%%%%%%%%%%%%%%%%%%%%%%%%%%%%%%%%%%%%%

In Table~\ref{thit}, the hits observed in the spectra of Q0307-195A,B
are reported together with the velocity differences between the
lines.  

To constrain cloud sizes, it can be assumed that a binomial random 
process produces the observed number of hits and misses; if the 
probability for a hit is given by $\Psi$, the likelihood function 
of this binomial process is 
${\cal L}({\cal N}_\mathrm{h}^o,{\cal N}_\mathrm{m}^o|\Psi)
=\Psi^{{\cal N}_\mathrm{h}^o}\,(1-\Psi)^{{\cal N}_\mathrm{m}^o}$ (FDCB).

$\Psi$ is the probability that both ray paths intersect 
the cloud, given that at least one ray path does.  
$\Psi$ is related to the probability $\Phi$, that one given LOS intersects 
the cloud given that the other adjacent ray path already does, by the 
formula:

\begin{equation}
 \Psi = \Phi/(2-\Phi).
\end{equation}

The functional form of the probability $\Phi$ depends on the 
assumptions for the geometrical shape of the absorbers. 
In the hypothesis of single radius, spherical clouds, with the 
definition $X=S/2\,R$, where $S$ is the proper separation of 
ray paths and $R$ is the cloud radius, it is found (McGill 1990):

\begin{equation}
\Phi = (2/\pi)\,[\arccos X - X(1-X^2)^{1/2}]\ \ \ \ \ \mathrm{ for}\ X<1,
\end{equation}

\noindent
and $\Phi=0$ otherwise.    

As a second model, clouds can be idealized as circular discs, with a given  
radius $R$, much larger then their thickness, and an observed 
inclination angle $\theta$. The probability that one ray path intersects 
the disc within the angle $\theta \rightarrow \theta + d\theta$, 
given that the other ray path already does, is  (McGill 1990):

\begin{eqnarray}
\Phi (\theta) &=& (\cos \theta/\pi)\bigg\{ \arccos \left(\frac{X}{\cos \theta}
\right) - \left(\frac{X}{\cos \theta}\right) \nonumber \\ 
&& \mbox{}  \times \left[ 1-\left(\frac{X}{\cos \theta}\right)^2
\right]^{1/2}\bigg\} \ \ \ \ \mathrm{ for}\ X<\cos \theta,
\end{eqnarray}

\noindent
and $\Phi=0$ otherwise.   
By integrating over $\theta$ for randomly oriented discs, the 
probability is then given by:

\begin{equation}
\Phi = \int^{\pi/2}_{-\pi/2} \Phi(\theta)\,d\theta.
\end{equation}

\vskip 12pt

In practice, from the observations we determine 
the numbers of hits and misses, 
${\cal N}_\mathrm{h}^o$ and ${\cal N}_\mathrm{m}^o$, 
and then we correct them for the accidental hits 
due to random velocity matches. 

Two different procedures have been devised to compute the value of 
${\cal N}_\mathrm{h}^o$:
in the first, the velocity separation between all the line pairs 
is computed and, when a given line combines to give more than one hit, 
only the one with the smallest velocity 
separation is taken into account, and the others are neglected. 
If, for example, lines 1 and 2 of spectrum A combine with lines 1 and 2 
of spectrum B forming the couples 1-1, 1-2 and 2-2, and the couple 1-2 
has the smallest velocity separation, we count just one hit. 
The second method takes the minimum value between the ``A vs. B'' hits  
and the ``B vs. A'' hits.

The two methods turned out to yield the same result in all the 
investigated cases. 

The number of observed misses is given by the sum of the two 
${\cal N}_\mathrm{m}^o$'s, ``A vs. B'' and ``B vs. A''.

To take into account the possibility of {\em accidental hits} due 
to random velocity matches, we have used simulations of Poissonian 
distributions of absorption lines in the two QSO LOSs, with a number of 
lines in each spectrum equal to the observed one.

For each simulation we have computed:
\begin{description}
\item[-] the number of accidental hits, ${\cal N}_\mathrm{r}$, 
applying the same procedures used for the real spectra; 
\item[-] the corrected number of hits, ${\cal N}_\mathrm{h}$, 
(misses, ${\cal N}_\mathrm{m}$) by 
subtracting ${\cal N}_\mathrm{r}$ from (adding $2\,{\cal N}_\mathrm{r}$ to)
the observed value.
\end{description}

Eventually, the probability distribution of the number of corrected hits 
${\cal C}({\cal N}_\mathrm{h})$ was evaluated. 
The probability ${\cal C}(0)$ takes into 
account all the occurrences with 
${\cal N}_\mathrm{r} \ge {\cal N}_\mathrm{h}^o$. 

\begin{figure}
\epsfxsize=90truemm
\epsffile{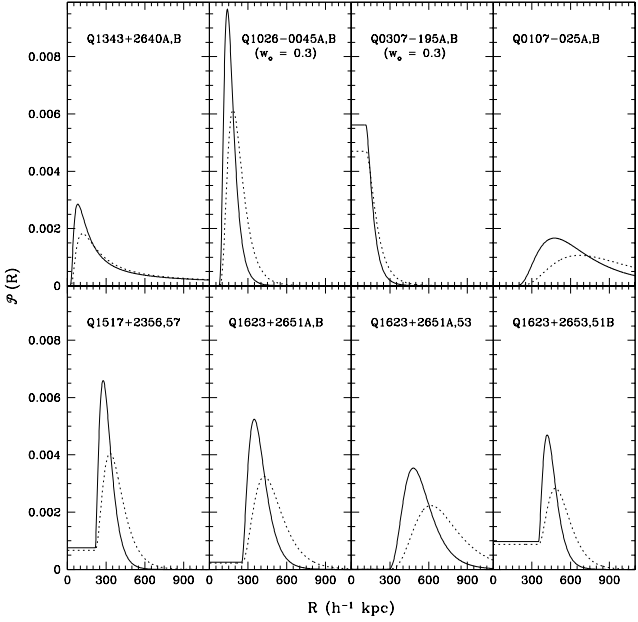}
\caption[]{\label{pdis}Probability distribution ${\cal P}(R)$ as given 
by eq. (8) for spherical absorbers ({\em solid curve}) and randomly 
inclined discs ({\em dotted curve}) as functions of cloud radius for
all the QSO pairs considered in the paper. 
The plateau at separation below $S/2$
arises because there is a non-vanishing probability that the number of 
corrected hits be 0 (${\cal C}(0)\ne 0$)}
\end{figure}

For each pair of values of corrected hits and misses, given 
the expressions for the probability distributions $\Phi$ and $\Psi$, 
it is possible to apply Bayes's Theorem (Press 1989, FDCB) to yield 
the a posteriori probability density for the radius $R$, 

\begin{equation}
{\cal P}(R,{\cal N_\mathrm{h}},{\cal N_\mathrm{m}}) = \frac{ {\cal L}
({\cal N_\mathrm{h}},{\cal N_\mathrm{m}}|\Psi)\, f(R)}
{\int_0^{\infty} {\cal L}({\cal N_\mathrm{h}},{\cal N_\mathrm{m}}|\Psi)\, 
f(R)\, dR}.
\end{equation}

At variance with FDCB, we have chosen as a prior distribution, $f(R)$,  
a uniform distribution for $R$ between 0 and $R_\mathrm{ max}$, and not 
a uniform distribution of $\Psi(R)$. 

It is worth stressing that a uniform prior distribution for the probability 
$\Psi$ introduces a cutoff at high and low $R$ that forces the 95 \% 
confidence intervals to be narrow also in situations with a very low number 
of hits or misses. 
%%%%%%%%%%%%%%%%%%% modified %%%%%%%%%%%%%%%%%%%%%%%%%%%%%%%%%%%%%%%%%%%
In particular, a spurious dependence of $R$ on the proper separation 
$S$ is induced in the case of very few hits, which is the common
situation at large separations,
%%%***
because the probability $\Psi(R)$ actually depends on $S$.   
%%%%%%%%%%%%%%%%%%%%%%%%%%%%%%%%%%%%%%%%%%%%%%%%%%%%%%%%%%%%%%%%%%%%%%%%
Since there is no particular justification to assume a uniform $\Psi(R)$, 
we have chosen a uniform distribution for $R$ which does not introduce 
biases in the results. 

\begin{figure}
\epsfxsize=90truemm
\epsffile{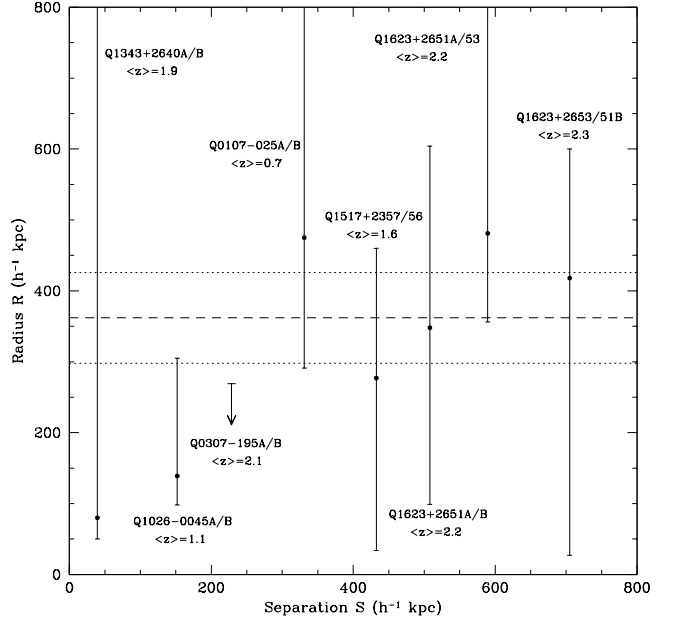}
\caption[]{\label{seprad:sph}Inferred cloud radius $R$ (and 95 \%
confidence interval) from a model assuming single-radius, unclustered 
spherical clouds, as a function of QSO pair sightline separation $S$. 
The dashed line shows the most probable value for $R$ and the dotted lines 
the 95 \% confidence interval (see text)}  
\end{figure}

\begin{figure}
\epsfxsize=90truemm
\epsffile{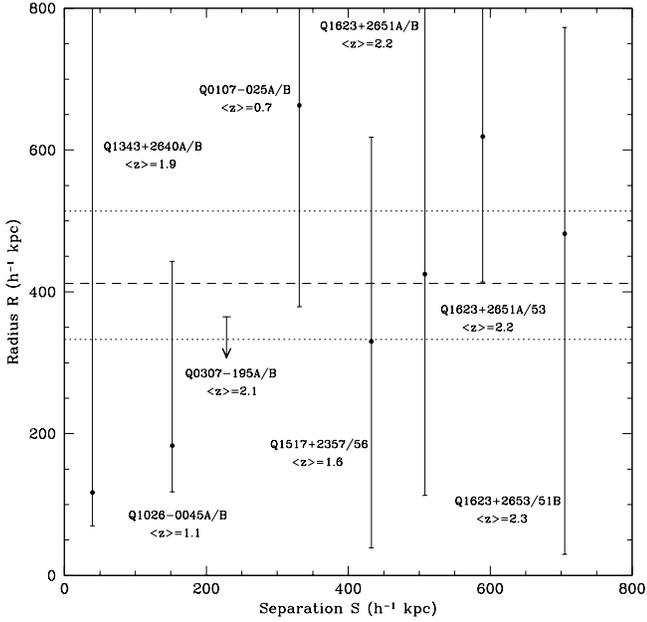}
\caption[]{\label{seprad:dsk}Same as Fig.~\ref{seprad:sph} but for 
single-radius, unclustered thin discs}
\end{figure}

Equation (5) then becomes

\begin{equation}
{\cal P}(R,{\cal N_\mathrm{h}},{\cal N_\mathrm{m}}) = \frac{ {\cal L}
({\cal N_\mathrm{h}},{\cal N_\mathrm{m}}|\Psi)}
{\int_0^{R_\mathrm{max}} {\cal L}({\cal N_\mathrm{h}},
{\cal N_\mathrm{m}}|\Psi)\, dR}.
\end{equation}

We have adopted as $R_\mathrm{max}$ a value of $3\ h^{-1}$ Mpc which 
represent  
the {\em mean free path} between two absorbers along the line of sight, as  
obtained from the line distribution $\partial^2n/\partial z\,\partial 
N_\mathrm{\ion{H}{i}}$ (Giallongo et al. 1996) at a redshift $z\sim2$, 
in the column density range $14\le \log N_\mathrm{\ion{H}{i}}\le 17$.
  
In the case of discs, the maximum value is increased by a factor 
$\sqrt 2$, yielding a value $R_\mathrm{max} \simeq 4 h^{-1}$ Mpc.

It has to be noted that the value of this maximum radius does not 
influence in practice the resulting modal radii and confidence intervals, 
because the probability distributions for $R$ usually go to zero well 
before this upper limit. $R_\mathrm{max}$ plays a significant role only if  
the number of misses is extremely low (1 or 0).

In the case of no hits, recipe (6) gives an incorrect result,  
in fact, the probability $\Phi$ becomes zero for $R < S/2$  by definition. 
In this extreme occurrence we have then to write:  

\begin{equation}
{\cal P}(R,{\cal N}_\mathrm{h},{\cal N}_\mathrm{m}) = A\times\left\{
        \begin{array}{ll}
         1 & \mbox{if $R < S/2$} \\
        {\cal L}({\cal N}_\mathrm{h},{\cal N}_\mathrm{m}|\Psi) & 
         \mbox{if $R\ge S/2$}
        \end{array}  
	\right.
\end{equation}

\noindent
where, $A$ is the proper normalization factor 
%%%***
which makes  
$\int_0^{R_\mathrm{max}}{\cal P}(R,{\cal N}_\mathrm{h},{\cal
N}_\mathrm{m}) =1$.

The global ${\cal P}(R)$ for a given QSO pair is finally obtained 
by the sum:

\begin{equation}
{\cal P}(R) = \sum {\cal P}(R,{\cal N}_\mathrm{h},{\cal N}_\mathrm{m}) 
\times {\cal C}({\cal N}_\mathrm{h}).
\end{equation}

\vskip 12pt

In order to investigate the possible correlations of the radius $R$ with 
the LOS separation and with the redshift, we have applied our analysis 
to 7 other data samples of comparable quality, published in the 
literature: Q1343+2640A,B (Crotts et al. 1994), Q1026-0045A,B
(Petitjean et al. 1998), Q0107-025A,B (D97), Q1517+2356,1517+2357 
(CF97), Q1623+2651A,1623+2653,1623+2651B (CF97).
%%%%%%%%%%%%%%%%%%%%%%% added part %%%%%%%%%%%%%%%%%%%%%%%%%%%%%%%%%%%% 
The QSO pair Q1343+2640A,B has been observed also by Dinshaw et
al. (\cite{dinshaw94}), 
%%%***
we have chosen to use Crotts et
al. (\cite{crotts94}) line list because of the larger wavelength
coverage and larger number of identified lines.  
%%%%%%%%%%%%%%%%%%%%%%%%%%%%%%%%%%%%%%%%%%%%%%%%%%%%%%%%%%%%%%%%%%%%%%

The relevant information are displayed in Table~\ref{t3}. 
Columns 6 and 7 report the observed number of hits and misses, 
respectively. 

The most probable number of accidental hits for each QSO pair is: 0 for 
1343+2640A,B; 0 for 1026-0045A,B (for both $W_o=0.2$ and $W_o=0.3$); 
2 for 0307-195A,B ($W_o=0.3$); 6 for 0307-195A,B ($W_o=0.2$); 
0 for 0107-025A,B; 1 for 1517+2356/57; 2 for 1623+2651A,B; 2 for 
1623+2651A/53 and 5 for 1623+2653,51B.  
 
In the actual computation of the cloud sizes the average value of the 
proper separation has been adopted. 

%%%%%%%%%%%%%%%%modified part%%%%%%%%%%%%%%%%%%%%%%%%%%%%%%%% 
In Fig.~\ref{pdis} the probability distributions given by equation (8) for 
all the QSO pairs presented in the paper are plotted for the two cases of
spherical and disc geometry.    
%%%%%%%%%%%%%%%%%%%%%%%%%%%%%%%%%%%%%%%%%%%%%%%%%%%%%%%%%%%%%%%%%%%

In Fig.~\ref{seprad:sph} and Fig~\ref{seprad:dsk} the values estimated for 
the modal radii are plotted as a function of $S$.
We 
%%%***
consider that if the {\em typical} radius  $R_{50} \le S/2$, 
where $R_{50}$ is defined as 
\begin{equation}
\int_0^{R_{50}}{\cal P}(R)= 0.5,
\end{equation}
only an upper limit can be set to the dimension of the absorber. 
This is the case for Q0307-195A,B (with $W_o \ge 0.3$ \AA) for 
which $R_{50} \simeq 90\ h^{-1}\ \mathrm{ kpc} < 114\ h^{-1}$ kpc for 
the spherical model and $R_{50} \simeq 107\ h^{-1}\ \mathrm{ kpc} < 
114\ h^{-1}$ kpc for the disc model.  
The dashed line represents the most probable modal radius,
$R = 362\ h^{-1}$ kpc, with 95\% confidence 
limits $298<R<426\ h^{-1}$ kpc  (spherical absorbers)
and $R = 412\ h^{-1}$ kpc, with 95\% confidence limits 
$333<R<514\ h^{-1}$ kpc (disc).
These values correspond to the peak of the total probability density 
obtained as the product of  the probability densities of the considered 
QSO pairs.  

A plot of the modal radius as a function of the average redshift of 
the \Lya\ forest of the corresponding QSO pair does not show any 
correlation between the two, at variance with the result obtained 
by Dinshaw et al. (\cite{dinshaw98}). 
 
The quality of the present data on the pair Q0307-195A,B and of the data by 
Petitjean et al. (\cite{ppjean}) for the pair Q1026-0045A,B allows defining 
a complete sample of lines down to an equivalent width $W_o = 0.2$ \AA.
The resulting number of hits and misses, the radius and confidence interval 
are reported in Table~\ref{t3}. 
%%The confidence interval is not compatible 
%%with the other intervals in Figs.~\ref{seprad:sph} and \ref{seprad:dsk}. 
The corresponding values of the global 
modal radius and 95 \% confidence interval are $R = 350\ h^{-1}$ kpc and 
$276 < R < 417\ h^{-1}$ kpc (sphere), $R = 408\ h^{-1}$ kpc and 
$325 < R < 501\ h^{-1}$ kpc (disc).  

Until fairly recently it has been conjectured that the typical
transverse sizes of the \Lya\ were of the order of a few tens of kpc.
The present analysis indicates about one order of magnitude larger
dimensions at $z \sim 2$, implying correspondingly larger ionizations
and masses.  With the presently inferred sizes and the typical UV
background conditions at $z\sim 2$ (Giallongo et al. 1996), optically
thin absorbers of $\log (N_{\ion{H}{i}}) \sim 14-15$ are expected to show a
remarkably small fraction of neutral Hydrogen, roughly $10^{-5.5}-10^{-5}$.  
In this way the mean intergalactic density contributed by
these clouds (Hu et al. 1995), especially in the spherical case, is
close to conflict with the baryon limit from nucleosynthesis $\Omega_b
\leq 0.015 h^{-2}$ (Walker et al. 1991). This potential incongruity
can however be alleviated by assuming a disc or other geometries.

\begin{figure*}
\epsfxsize=17cm
\epsffile{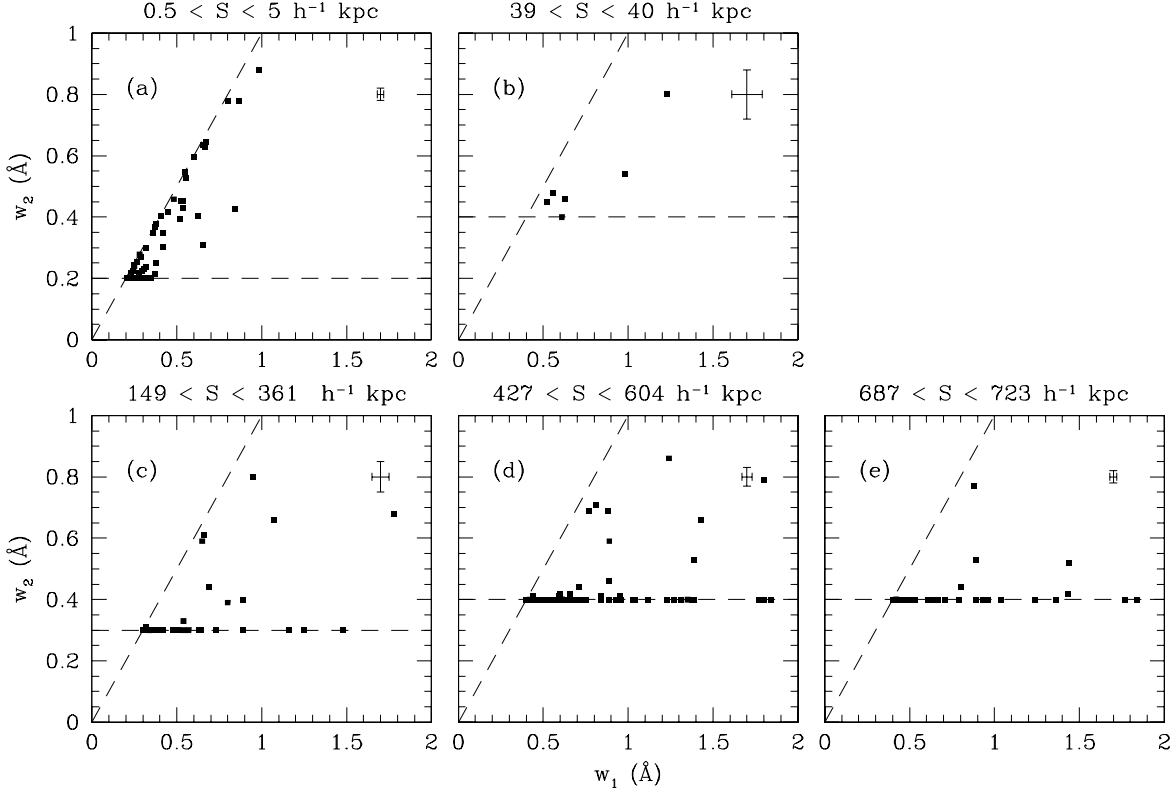}
\vskip -6cm
\caption[]{\label{eqwdt}Equivalent widths of the hits and misses for the 
8 QSO pairs 
in our sample plus the data by Smette et al. (1995), they are grouped 
as follows: (a) HE1104-1805; (b) Q1343+2640A,B; (c) Q1026-0045A,B 
Q0307-195A,B and Q0107-025A,B; (d) Q1517+2356,1517+2357  Q1623+2651A,B and 
Q1623+2651A,1623+2653; (e) Q1623+2653,1623+2651B. 
The equivalent widths are arranged such that $W_1 \equiv \mathrm{ max}
(W_A,W_B)$ and $W_2 \equiv \mathrm{ min}(W_A,W_B)$. 
The horizontal dashed lines represent the rest equivalent width threshold: 
$W_o >0.2$ for (a), $W_o >0.3$ for (b), (c) and 
$W_o >0.4$ for (d), (e). The error bars in the upper right corners of the plots 
represents the average $1\ \sigma$ errors for the equivalent width 
}  
\end{figure*}

\subsection{Towards more realistic models}

The approach described in the previous section, although customarily
used to 
determine the size of the \Lya\ absorbers, is expected to give just a rough 
idea of the dimension of these structures, due to the simplistic hypotheses 
made on their geometry. 

In particular, the information on the equivalent width of the lines are not 
taken into account, since no density distribution is assumed inside the 
absorbers. Besides, all the clouds are assumed to have the same radius. 

%%%%%%%%%%%%%%%%%%%%%%%%modified part%%%%%%%%%%%%%%%%%%%%%%%%%%%%%%
In the literature (Smette et al. 1992; Smette et al. 1995; Charlton
et al. 1997; Dinshaw et al. 1997) this problem has
been tackled  
adopting a statistical technique that utilizes the information about 
the equivalent width of the lines.
%%%%%%%%%%%%%%%%%%%%%%%%%%%%%%%%%%%%%%%%%%%%%%%%%%%%%%%%%%%%%%%%%%%%%

Dinshaw et al. (1997)
adopted a power-law column density profile for the absorbers, $N(r) = 
N_\mathrm{lim} (r/R_0)^{-\gamma}$, where $N_\mathrm{lim}=1.26\times 
10^{14}$ cm$^{-2}$ for $b=35$ km s$^{-1}$ is the limiting 
column density of the sample, $R_0$ is the radius of the absorber at $N = 
N_\mathrm{lim}$ and $\gamma = 4$, and 
three simple geometries with fixed radius: spheres, randomly inclined 
discs, and pseudo-filamentary structures (approximated as discs with fixed 
inclination $\cos i = 0.2$).  

The column density distribution along paired lines of sight for each 
of the geometric models considered is simulated by means of the Monte Carlo 
techniques outlined by Smette et al. (\cite{smette92}). 
This distribution can be converted into the corresponding equivalent width 
distribution adopting an approximation to the curve of growth (Chernomordik 
\& Ozernoy 1993).

In Fig.~\ref{eqwdt} the equivalent widths of the hits and misses are 
plotted for an enlarged sample of QSO pairs formed by our previously 
described 8 cases plus the lensed QSO HE 1104-1805A,B observed by 
Smette et al. (\cite{smette95}). 
The pairs are grouped according to their typical proper separation. 
%%%%%%%%%%%%%%%%%%%%%%%%%added part%%%%%%%%%%%%%%%%%%%%%%%%%%%%%%%%%
It should be noted that
it is now certain that the double QSO HE 1104-1805A,B is actually a
gravitational lens system, as the lensing galaxy has been detected
(Courbin et al. 1998; Remy et al. 1998). 
%% As a consequence, the separation between the two light beams
%% is much smaller than the one given in Fig.~\ref{eqwdt} and decreases quickly
%% as the redshift of the clouds gets close to the quasar emission
%% redshift.
%%%***
The exact values of the separation between the two light beams depend on the 
still badly determined lens redshift. The values we reported in 
Fig.~\ref{eqwdt} have been estimated using  Fig.~6 of the paper by 
Smette et al. (\cite{smette95}), for the redshift $z_{\rm lens}=1.32$,  
favoured by Remy et al. (1998). 
%%%%%%%%%%%%%%%%%%%%%%%%%%%%%%%%%%%%%%%%%%%%%%%%%%%%%%%%%%%%%%%%%%%%%

It is clear from the figure that the fraction of hits decreases as
the proper separation increases, while, at the same time, 
the correlation between the equivalent widths of the absorbers in the two 
LOS becomes poorer and poorer.  
 
The present observations can be compared with the simple model predictions 
of D97 (their Fig.~12). In particular, our Fig.~\ref{eqwdt}(a) shows 
reasonable agreement with the prediction of a disc model with 
$R_0 \sim 7\,S$ 
that is $R_0 \sim 90\ h^{-1}$ kpc; panel (b) agrees with the case $R_0
= 2-3\ S$  
for both filament and disc model, that is $R_0 \simeq 80-120\ h^{-1}$ kpc; 
panel (c) can be explained by filaments and disc with $R_0 = S$ which 
implies $R_0 \sim 220\ h^{-1}$ kpc and, finally, panels (d) and (e) 
agrees with filament and disc models with $R_0 \le S$ or $R_0 \lsim 
500-700\ h^{-1}$ kpc. 
Altogether a disc geometry with $R_0 \sim 100-200 \ h^{-1}$ kpc
seems to be favored.

It should be noted, however, that the number of degrees of freedom is not 
the same in the three models: once a given radius is chosen in the disc 
idealization the two orientation angles can vary at random, for the 
filament geometry one angle is fixed, and for the sphere no angular 
variable exists to define an orientation. 
In this way, it is obviously easier for the disc model to reproduce the 
real data that show a substantial amount of scatter due to measurement 
errors and possibly to a non-unique value of the cloud size.   

In fact, by introducing one more degree of freedom in the spherical 
cloud model: i.e. allowing the radius $R$ to vary according to one of 
the two distribution: 
(1)exponential, and (2)power law with indices $\alpha = -4, -3,..., 5, 6$, 
it is found that a power law distribution with $\alpha=3$ provides an 
equally reasonable representation of the observed equivalent width 
distribution (D97).

\section{Conclusions}

%%%%%%
We have presented new spectra of the quasar pair Q0307-195A,B. 

The number of detected lines is almost doubled with respect to the 
previous study of this pair (Shaver \& Robertson 1983). 
One new metal system has been identified in object A on the basis of the  
\ion{C}{iv} doublet and one possible system is found also in object B. 

In order to study the shape and dimension of the \Lya\ absorbers, we 
have added
%%%***
to the observations of the present QSO pair, 
data of comparable resolution taken from the literature to 
create a sample of 8 QSO pairs.  
We looked for hits and misses in the pair spectra and 
computed the probability distribution for a given radius $R$ in the case of 
spherical and disc geometry basing our statistical analysis on the approach 
outlined by Fang et al. (\cite{fang96}).  
We modified some of their assumptions in order to deal correctly 
with the extreme cases of very low numbers (1 or 0) of hits or misses and 
avoid any spurious dependence on the separation between 
lines of sight.
 
Our results can be summarized as follows:

\begin{description}
\item[-] the total probability density obtained as the product of the 
probability densities of the considered QSO pairs, gives a modal radius  
$R = 362\ h^{-1}$ kpc, with 95\% confidence 
limits $298<R<426\ h^{-1}$ kpc (spherical absorbers)  
and $R = 412\ h^{-1}$ kpc, with 95\% confidence limits 
$333<R<514\ h^{-1}$ kpc (disc).

\item[-] No significant correlation is detected between the modal
radius and the mean proper separation of the pair, nor between the
modal radius and the mean redshift of the \Lya\ forest region. 

\item[-] The comparison of the distribution of the equivalent widths of the 
observed coincident lines with the predictions of Monte Carlo
simulations for three simple geometrical models, spheres, discs and
filaments with a power-law column density profile $N(r) \propto
(r/R_0)^{-\gamma}$ (Dinshaw et al. 1997), seems to indicate a disc
shape with $R_0 \sim 100-200\ h^{-1}$ kpc and $\gamma \sim 4$ as the
most suitable.  This could however be simply the result of the
different number of parameters in the three models.  If, for example,
a further degree of freedom is allowed to the model based on spheres
in the form of a power-law distribution of radii, a satisfactory
consistency with the data is obtained.
 
\end{description}

\begin{acknowledgements}
We are grateful to J. Bergeron for enlightening suggestions.
%%%*** THANKS TO THE REFEREE
We thank the referee, Dr. A. Smette, for the stimulating  
comments.  
SC acknowledges the support of the ASI contract ARS-96-176 and 
of the Formation and Evolution of Galaxies
network set up by the European Commission under contract ERB FMRX-CT96-086
of its TMR programme.
\end{acknowledgements}

\end{document}